\documentclass[
reprint,
bibnotes,
amsmath,
amssymb,
amsfonts,
floatfix,
showkeys,
showpacs,
aps,
prx
]{revtex4-2}

\usepackage{graphicx}
\usepackage[version=3]{mhchem} % Formula subscripts using \ce{}
\usepackage{bm}
\usepackage{chemformula} 
\usepackage{multirow}
\usepackage{color,soul,xcolor}

\begin{document}

%%%%%%%%%%%%%%%%%%%%%%%%%%%%%%%%%%%%%%%%%%%%%%%%%%%%%%%%%%%%%%%%%%%%%
\title{The Resistive Nature of Decomposing Interfaces  \\
of Solid Electrolytes with  Alkali Metal Electrodes}

\author{Juefan Wang}
\affiliation{Department of Materials Science and Engineering, National University of Singapore, 9 Engineering Drive 1, 117575, Singapore}

\author{Abhishek A. Panchal}
\affiliation{Department of Materials Science and Engineering, National University of Singapore, 9 Engineering Drive 1, 117575, Singapore}

\author{Gopalakrishnan Sai Gautam}
\affiliation{Department of Materials Engineering, Indian Institute of Science, Bangalore 560012, India}

\author{Pieremanuele Canepa}
\affiliation{Department of Materials Science and Engineering, National University of Singapore, 9 Engineering Drive 1, 117575, Singapore}
\affiliation{Department of Chemical and Biomolecular Engineering, National University of Singapore, 4 Engineering Drive 4, 117585, Singapore}
\email{pcanepa@nus.edu.sg}

%%%%%%%%%%%%%%%%%%%%%%%%%%%%%%%%%%%%%%%%%%%%%%%%%%%%%%%%%%%%%%%%%%%%%

%%%%%%%%%%%%%%%%%%%%%%%%%%%%%%%%%%%%%%%%%%%%%%%%%%%%%%%%%%%%%%%%%%%%%
\begin{abstract}
A crucial ingredient in lithium (Li) and sodium (Na)-ion batteries (LIBs and NIBs) is the  electrolytes. The use of Li-metal (Na-metal) as anode in liquid electrolyte LIBs (NIBs) is constrained by several issues including thermal runway and flammability, electrolyte leakage, and limited chemical stability.  Considerable effort has been devoted toward the development of solid electrolytes (SEs) and all-solid-state batteries, which are presumed to mitigate some of the issues of Li-metal(Na-metal) in contact with flammable liquid electrolytes. However, most SEs, such as \ch{Li3PS4}, \ch{Li6PS5Cl} and \ch{Na3PS4} readily decompose against the highly reducing Li-metal and Na-metal anodes. Using first-principles calculations we elucidate the stability of more than 20 solid$||$solid interfaces formed between the decomposition products of \ch{Li3PS4}, \ch{Li6PS5Cl} (and \ch{Na3PS4}) against the Li-metal (Na-metal) electrode. We suggest that the work of adhesion needed to form a hetereogenous interfaces is an important descriptor to quantify the  stability of interfaces. Subsequently, we clarify the atomistic origins of the resistance to Li-ion transport at interfaces of the Li-metal anode and selected decomposition products  (\ch{Li3P}, \ch{Li2S} and \ch{LiCl}) of SEs, via a high-fidelity machine learned potential (MLP). Utilising an MLP enables nano-second-long molecular dynamics simulations on `large' interface models (here with 8320 atoms), but with similar accuracy to first-principles approaches.  Our simulations demonstrate that  the interfaces formed between Li-metal and argyrodite (e.g., {\ch{Li6PS5Cl}}) decomposition products are resistive to Li-ion transport. The implications of this study are important since binary compounds  are commonly found in the vicinity of Li(Na)-metal upon chemical and/or electrochemical decomposition of ternary and quaternary SEs.
\end{abstract}

%%%%%%%%%%%%%%%%%%%%%%%%%%%%%%%%%%%%%%%%%%%%%%%%%%%%%%%%%%%%%%%%%%%%%
\maketitle

%%%%%%%%%%%%%%%%%%%%%%%%%%%%%%%%%%%%%%%%%%%%%%%%%%%%%%%%%%%%%%%%%%%%% %
%Start the main part of the manuscript here.
%%%%%%%%%%%%%%%%%%%%%%%%%%%%%%%%%%%%%%%%%%%%%%%%%%%%%%%%%%%%%%%%%%%%%
\section{Introduction}
%%%%%%%%%%%%%%%%%%%%%%%%%%%%%%%%%%%%%%%%%%%%%%%%%%%%%%%%%%%%%%%%%%%%%

\noindent  Rechargeable lithium-ion batteries (LIBs) keep gaining importance for the development of the next-generation energy storage devices and electric vehicles because of their outstanding gravimetric and volumetric energy densities.\cite{goodenoughChallengesRechargeableLi2010,xuLithiumMetalAnodes2014,linRevivingLithiumMetal2017,krauskopfPhysicochemicalConceptsLithium2020,chengSafeLithiumMetal2017}   Lithium metal batteries (LMBs) utilizing Li-metal anodes ---that can achieve unprecedented  energy  densities theoretically, as compared to LIBs--- have become one of the central topics of current research in rechargeable batteries. \cite{chengSafeLithiumMetal2017,fangPressuretailoredLithiumDeposition2021,krauskopfPhysicochemicalConceptsLithium2020}  The primary challenge in constructing practical LMBs is stabilizing the Li-metal$||$electrolyte interface, with scientific studies mostly focused on identifying  electrolyte formulations with limited reactivity and/or suitable additives.\cite{goodenoughChallengesRechargeableLi2010,xuNonaqueousLiquidElectrolytes2004,chengSafeLithiumMetal2017} Stabilizing the metal$||$electrolyte interface is also a bottleneck in developing Na-metal batteries (NMBs).\cite{yabuuchiResearchDevelopmentSodiumIon2014,nayakLithiumIonSodiumIonBatteries2018,delmasSodiumSodiumIonBatteries2018,chengSafeLithiumMetal2017, krauskopfPhysicochemicalConceptsLithium2020} 

Solid electrolytes (SEs) are critical components in the development of LMBs and solid-state LIBs.\cite{haruyamaSpaceChargeLayer2014,tangProbingSolidSolid2018,wuNewInsightsInterphase2018,wenzelInterfacialReactivityInterphase2018,richardsInterfaceStabilitySolidState2016,gaoFirstPrinciplesStudyMicroscopic2021,famprikisFundamentalsInorganicSolidstate2019,famprikisPressureMechanochemicalEffects2020} Besides acting as separators between electrodes, SEs are also expected to alleviate some of the safety issues between Li-metal anodes and liquid electrolytes. \cite{baggettoHighEnergyDensity2008,chengSafeLithiumMetal2017,krauskopfPhysicochemicalConceptsLithium2020} Nevertheless, numerous reports have demonstrated high electrochemical instabilities of SEs when in contact with Li-metal anode (and other electrode materials).\cite{krauskopfPhysicochemicalConceptsLithium2020}  For example, sulfur-containing SEs are unstable against Li-metal, resulting in the formation of undesired decomposition products, which may resist Li-ion transport and/or facilitate electron transport.\cite{richardsInterfaceStabilitySolidState2016,wenzelInterfacialReactivityInterphase2018,wuNewInsightsInterphase2018,tangProbingSolidSolid2018,zhuFirstPrinciplesStudy2016,lacivitaInitioInvestigationStability2019,schwietertClarifyingRelationshipRedox2020}  Thus, the stabilization of interfaces formed between Li-metal (or other alkali-metal electrodes) and SEs remains a significant bottleneck in designing practical solid-state batteries.

Electrolyte decomposition occurs at small length scales away from the exteriors of the cell packs that constitute a battery. Therefore, the characterization of decomposition products in fully assembled and operating devices requires dedicated custom-made and expensive tools.\cite{wenzelInterfacialReactivityInterphase2018,wuNewInsightsInterphase2018,hoodElucidatingInterfacialStability2021,chengUnveilingStableNature2020} A number of reports have analyzed the compositions, structures, and formation mechanisms of the decomposing products of SEs against metal electrodes( metal electrode$||$SE).\cite{wenzelInterfacialReactivityInterphase2018,wuNewInsightsInterphase2018,tangProbingSolidSolid2018,richardsInterfaceStabilitySolidState2016,zhuFirstPrinciplesStudy2016,lacivitaInitioInvestigationStability2019,schwietertClarifyingRelationshipRedox2020} For example, X-ray photoemission spectroscopy (XPS) experiments by \citeauthor{wenzelInterfacialReactivityInterphase2018}\cite{wenzelInterfacialReactivityInterphase2018} reported that \ch{Li6PS5X} (with X=\ch{Cl}, \ch{Br} and \ch{I}), upon contact with Li-metal, forms \ch{Li2S}, \ch{LiX}, and \ch{Li3P}. As a result, the decomposition products of metal electrode$||$SE are expected to be multiphased and highly heterogeneous, which complicates the description of ionic transport across interfaces. Furthermore, the structures and properties of the metal electrode$||$SE interfaces are expected to be markedly different from the bulk materials. A detailed study of the interfacial properties, particularly ionic transport, is needed for the advancement of solid-state batteries.  
 
Another aspect of solid-state batteries relates to the mechanical stability (i.e., adhesion) of the solid$||$solid interfaces that are electrochemically formed. The loss of contact due to lack of adhesion between  Li-metal and SEs appears as a major cause driving the buildup of interfacial impedance in solid-state devices.\cite{krauskopfPhysicochemicalConceptsLithium2020,chengSafeLithiumMetal2017,yuUnravellingLiIonTransport2016} To evaluate the mechanical stabilities of the interfaces, \citeauthor{lepleyModelingInterfacesSolids2015}\cite{lepleyModelingInterfacesSolids2015} have performed  accurate  first-principles calculations of several Li-metal$||$SE interfaces (such as, Li$||$\ch{Li2O}, Li$||$\ch{Li2S}, Li$||$\ch{Li3PO4} and Li$||$\ch{Li3PS4}) and found that all interfaces were stable except \ch{Li}$||$\ch{Li3PS4}. Other studies have investigated the effects of  stability of heterogeneous interfaces on the Li-ion transport properties.\cite{yangMaintainingFlatLi2021,seymourSuppressingVoidFormation2021,yangInterfacialAtomisticMechanisms2021}

\citeauthor{yangMaintainingFlatLi2021}\cite{yangMaintainingFlatLi2021} have proposed that an interface with good adhesion, i.e."lithiophilic interface" can result in a faster critical stripping current density, which is crucial to prevent dendrite growth. Recently, \citeauthor{seymourSuppressingVoidFormation2021}\cite{seymourSuppressingVoidFormation2021} have shown that Li (or Na)-ion transport across alkali-metal$||$SE interfaces correlates directly with the interfacial adhesion.
\citeauthor{yangInterfacialAtomisticMechanisms2021}{\cite{yangInterfacialAtomisticMechanisms2021}} have employed classical molecular dynamics (MD) to study the process of Li plating and stripping on solid \ch{Li2O}, showing that a coherent interface with strong interfacial adhesion and fast Li-ion diffusion can prevent pore formation at the interface. Here, we perform a systematic investigation including a larger data set of solid$||$solid interfaces, particularly focusing on the correlation between the atomistic structure of interfaces and ionic transport, which is presently lacking.

Further, we address  the interfacial stability and Li-ion mobility of multiple interfaces formed  between Li-metal electrode and decomposition products of topical SEs, such as,  \ch{Li3PS4},\cite{mizunoNewHighlyIonConductive2005,hakariAllsolidstateLithiumBatteries2015,richardsInterfaceStabilitySolidState2016,lepleyModelingInterfacesSolids2015} argyrodite-\ch{Li6PS5Cl}\cite{wenzelInterfacialReactivityInterphase2018} and LiPON, with the general formula \ch{Li_xPO_yN_z}.\cite{hoodElucidatingInterfacialStability2021,chengUnveilingStableNature2020,schwobelInterfaceReactionsLiPON2015}  We also analyze the \ch{Na}$||$\ch{Na2S} and \ch{Na}$||$\ch{Na3P} interfaces, which form upon the decomposition of \ch{Na3PS4} against \ch{Na}-metal.\cite{wuNewInsightsInterphase2018} We perform large-scale MD simulations of selected interfaces,(i.e., Li$||$\ch{Li3P}, Li$||$\ch{Li2S} and Li$||$\ch{LiCl}) based on  high-fidelity machine learned potentials (MLPs) trained on accurate first-principles data, which carry the accuracy of \emph{ab initio}
molecular dynamics (AIMD) while give access to appreciably larger time and length scale simulations.

We reveal that the mechanical stabilities of the Li (or Na)-metal$||$SE interfaces are primarily governed by the atomistic structures of the interfaces, which in turn are dependent on the surface orientations and/or terminations of the decomposition products. Further, we show that the interfaces formed between Li-metal and decomposition products of the argyrodite \ch{Li6PS5Cl} SE (i.e., \ch{Li3P}, \ch{Li2S} and \ch{LiCl}) are resistive to Li-ion transport, explaining the observed impedance buildup. Our results provide insights to engineer solid$||$solid interfaces with better interfacial stability and improved ionic transport. 

%%%%%%%%%%%%%%%%%%%%%%%%%%%%%%%%%%%%%%%%%%%%%%%%%%%%%%%%%%%%%%%%%%%%%
\section{Construction of Interfaces of Decomposition Products and Metal Anodes}
\label{sec:quantities}
%%%%%%%%%%%%%%%%%%%%%%%%%%%%%%%%%%%%%%%%%%%%%%%%%%%%%%%%%%%%%%%%%%%%%
We discuss the procedure to build heterogeneous interfaces between an alkali-metal (Li or Na) with one of their binary compounds (e.g., \ch{Li3P}), formed as a result of SE decomposition. In constructing the heterogeneous interfaces between the alkali-metal (e.g., Li or Na) and the binary compounds, we  identify stable stoichiometric surfaces (following Tasker's criteria\cite{taskerStabilityIonicCrystal1979}) with low surface energies, $\gamma$, of both materials, which are paired into an interface (see Table S1 of the Electronic Supplementary Information, ESI).  To describe $\gamma$, we have used the slab model in Eq.~\ref{eq:gamma}.\cite{butlerDesigningInterfacesEnergy2019}
\begin{equation}
\label{eq:gamma}
\gamma = \lim _{N\rightarrow \infty} \frac{1}{2S} \left [ E_\mathrm {slab} ^N - N E_\mathrm{bulk} \right ] 
\end{equation}
where $S$ is the surface area of the slab, $E_\mathrm{slab}^N$ is the energy of the relaxed slab containing $N$ formula units, and $E_\mathrm{bulk}$ is the energy per formula unit of the bulk structure. The energies of Eq.~\ref{eq:gamma} (and the following equations) are Gibbs energies, which we  approximate by density functional theory (DFT, see Sec.~\ref{sec:methods}) total energies ignoring $pV$ and entropic contributions. Slab models included sufficient number of layers and a vacuum of 15 \AA{} to converge $\gamma$  to within $\pm$0.01~J~m$^{-2}$.

\begin{figure*}[!ht]
\includegraphics[width=\textwidth]{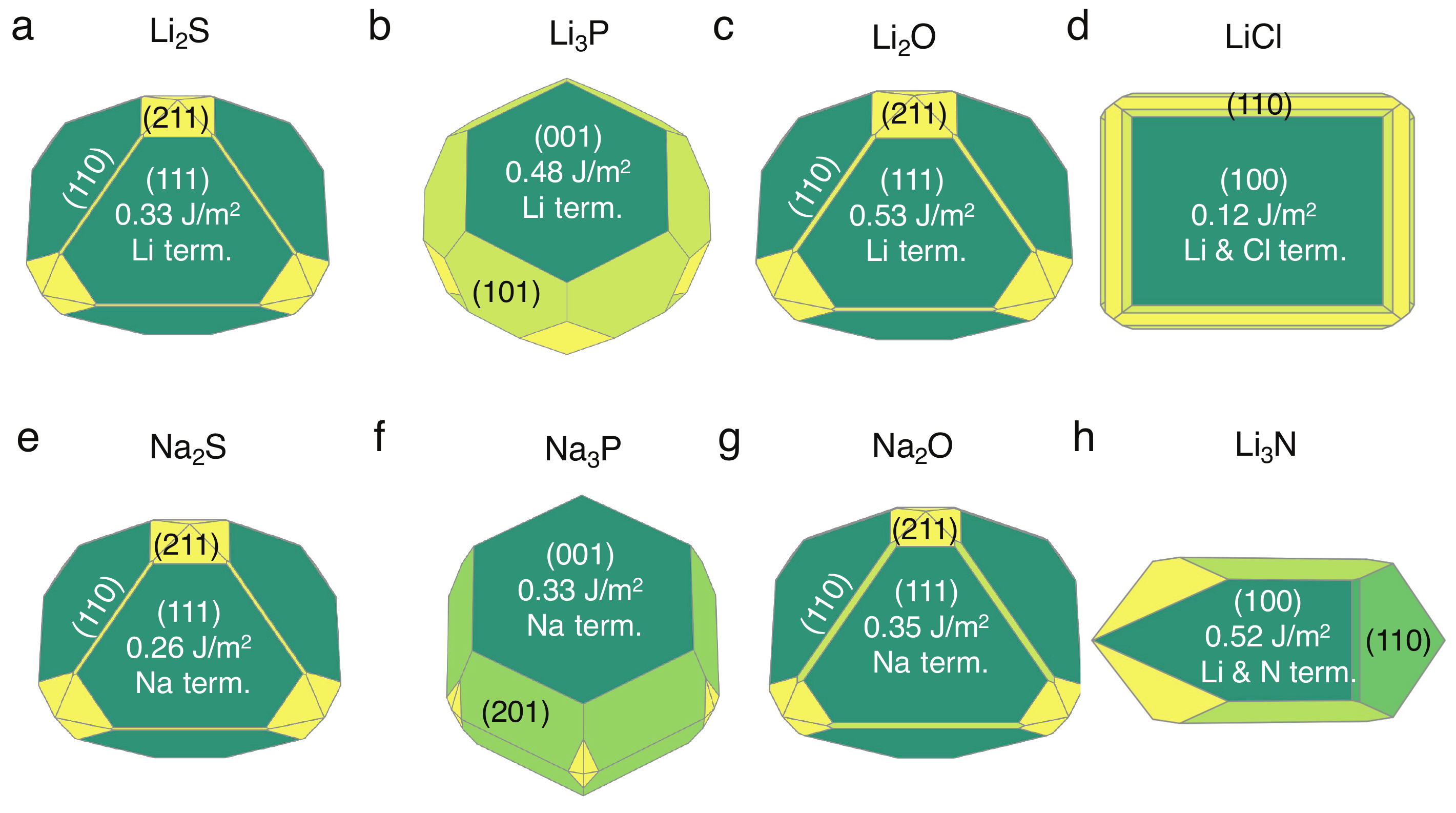}
\caption{
Computed Wulff shapes of binary compounds \ch{Li2S} (panel a), \ch{Li3P} (b), \ch{Li2O} (c), \ch{LiCl} (d), \ch{Na2S} (e), \ch{Na3P} (f), \ch{Na2O} (g), and \ch{Li3N} (h), with their corresponding surface energies (in J~m$^{-2}$). The chemical nature of the surface terminations (term.) are indicated. Wulff polygons are constructed using  stoichiometric, non-polar, and symmetric (including an inversion symmetry)  surfaces. 
}
\label{fig:wulff}
\end{figure*}

The set of stable surfaces in Li(or Na) metal and binary compounds considered, and their corresponding $\gamma$, are displayed in the Wulff shapes  of Figure~\ref{fig:wulff}.\cite{einsteinEquilibriumShapeCrystals2015, tranSurfaceEnergiesElemental2016} $\gamma$ values, not shown in Figure~{\ref{fig:wulff}}, are included in Table S2-S3 of ESI. The (100) surface of Li-metal has the lowest surface energy of $\sim$0.46 J~m$^{-2}$, while for Na-metal, the (100) and (110) surfaces have similar $\gamma$ values, $\sim$0.22 J~m$^{-2}$ and $\sim$0.21 J~m$^{-2}$. In \ch{Li2S}, \ch{Li2O}, \ch{Na2S}, and \ch{Na2O}, the (111) facet dominates the Wulff shape, while for \ch{Li3P}, \ch{LiCl}, \ch{Li3N}, and \ch{Na3P}, the \{001\}-type surfaces have the lowest $\gamma$ (Figure~\ref{fig:wulff}). Our calculated surface energies, $\sim$0.33 J~m$^{-2}$ for the (111) surface and $\sim$0.51 J~m$^{-2}$ for the (110) surface of \ch{Li2S}, as well as $\sim$0.53 J~m$^{-2}$ for the (111) surface of \ch{Li2O} are consistent with previous literatures.\cite{chenMetalizationLiParticle2014,mackrodtAtomisticSimulationSurfaces1989}   
\ch{Li3N} and \ch{LiCl} exhibit stable facets that have both Li and anion species, while other compounds have stable facets exposing a Li (or Na) layer. 

The  surfaces of Figure{~\ref{fig:wulff}} are subsequently paired to form heterogeneous interfaces. Different metrics serve to quantify the effect of mechanical strain and/or the chemical bond formation/destruction at the interface.\cite{liuInterfacialStudySolid2016, hashibonStructureAbruptCopper2005}  The interface formation energy ($E_f$ in Eq.~\ref{eq:interfaceforamation}) is the energy difference between the interface model and bulk structures of ${A}$ and ${B}$, and includes both mechanical (i.e., elastic strain) and chemical components.\cite{butlerDesigningInterfacesEnergy2019} 
\begin{equation}
\label{eq:interfaceforamation}
 E_f=\frac{E_{AB}- \left[ {N_AE}_A + {N_BE}_B\right]}{2S}
\end{equation}
where $S$ is the surface area of the interface, $E_{AB}$ is the energy of the fully relaxed interface model, containing $N_A$ and $N_B$ formula units of materials $A$ and $B$, whose bulk energies are $E_A$ and $E_B$. Elastic stress can arise in interfaces displaying large lattice mismatch, and "absorbed" by the interface through the release of the stress energy, via formation of  dislocations.{\cite{benedekEffectMisfitHeterophase2002,dalvernyInterfaceElectrochemistryConversion2011}} By removing the elastic strain from $E_f$ (of Eq.~{\ref{eq:interfaceforamation}}), we obtain two important descriptors: \emph{i}) the interfacial energy, $\sigma$ of Eq.~{\ref{eq:sigma}}, and \emph{ii}) the work of adhesion,  $W_\mathrm{adhesion}$ of Eq.~{\ref{eq:workadhesion}}, which are paramount in evaluating the overall stability of interfaces. $\sigma$ quantifies the formation (or destruction) of chemical bonds as the interface is created, excluding all mechanical contributions. 
\begin{equation}
\label{eq:sigma}
\sigma = \frac{E_{AB}-\left[{N_AE}_{A(z)}+{N_BE}_{B(z)}\right]}{2S}
\end{equation}
where $E_{A(z)}$ and $E_{B(z)}$ are the energy per formula unit of the bulk $A$ and $B$, as obtained from a constrained relaxation along the normal direction ($z$) to the interface, where the in-plane lattice vectors of the bulk structures are fixed to those of the fully relaxed interface. It follows that, the elastic strain energy associated with the interface is calculated as $E_f-{\sigma}$.  

The work of adhesion, $W_\mathrm{adhesion}$ (of  Eq.~\ref{eq:workadhesion}) is the work done to part two adherent surfaces to an infinite distance, and quantifies the mechanical stability of an interface.
\begin{equation}
\label{eq:workadhesion}
W_\mathrm{adhesion}=\gamma_A+\gamma_{B} - \sigma
\end{equation}
where $\gamma_A$ and $\gamma_B$ (Eq.~\ref{eq:gamma}) are the surface energies of materials $A$ and $B$. Nominally, small (positive) values of $\sigma$ and large (positive) values of $W_\mathrm{adhesion}$ are indicative of high interfacial stability. To account for the effect of elastic strain, Eq.~\ref{eq:workadhesion2} gives an alternative definition of $W_\mathrm{adhesion}$ 
\begin{equation}
\label{eq:workadhesion2}
W_\mathrm{adhesion}=\gamma_A+\gamma_{B} - E_f
\end{equation}
For creating interface models, we use the algorithm by \citeauthor{taylorARTEMISInitioRestructuring2020},\cite{taylorARTEMISInitioRestructuring2020}  which samples the  configurational space to find interface models that minimize the lattice mismatch between two materials. While pairing surfaces, we used the in-plane lattice constants of the binary compounds (e.g., \ch{Li3P}) and applied a lattice mismatch-induced strain to the metal surface, since the bulk moduli of binary compounds are typically greater than the alkali-metals (i.e., Li and Na).\cite{haruyamaSpaceChargeLayer2014,dengElasticPropertiesAlkali2016}  The constructed interface models are symmetric; for example, \ch{Li2S}$||$\ch{Li}-metal consists of two identical interfaces that forms a \ch{Li2S}$||$\ch{Li-metal}$||$\ch{Li2S} system, as displayed in panels c and d in Figure~\ref{fig:interfaces}. The slab thickness of binary compounds is typically $\sim$10~\AA, which is sufficient to distinguish the interface features from their bulk-like properties. However, thicker slabs are required for Li ($\sim$12~\AA) and Na ($\sim$14~\AA{}) to distinguish the interface regions from the bulk region. \cite{lepleyModelingInterfacesSolids2015} 

%%%%%%%%%%%%%%%%%%%%%%%%%%%%%%%%%%%%%%%%%%%%%%%%%%%%%%%%%%%%%%%%%%%%%
\section{Stability of Interfaces of Decomposition Products and Metal Electrodes}
%%%%%%%%%%%%%%%%%%%%%%%%%%%%%%%%%%%%%%%%%%%%%%%%%%%%%%%%%%%%%%%%%%%%%

\begin{figure*}[!ht]
\includegraphics[width=\textwidth]{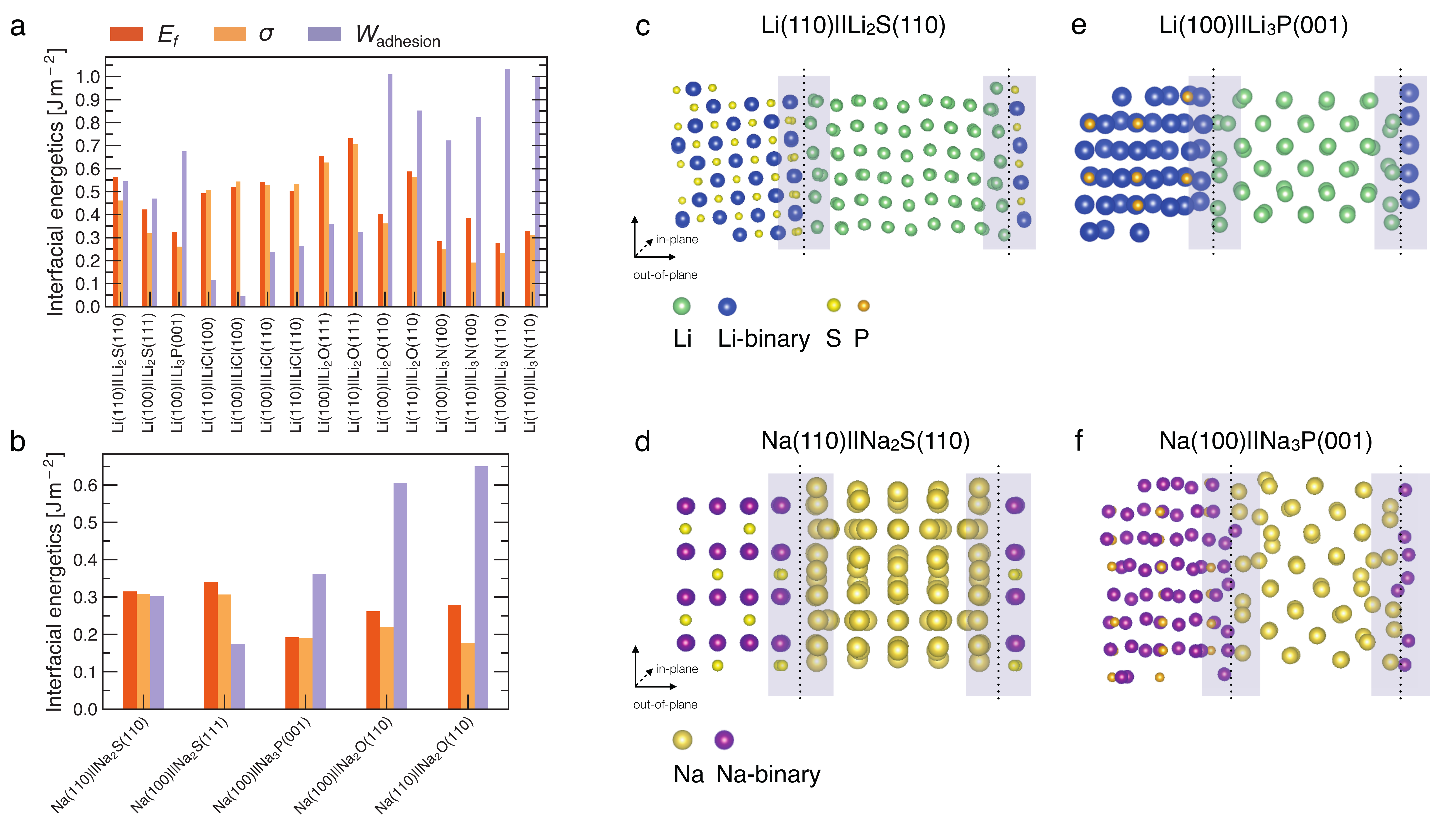}
\caption{
 Computed interfacial quantities (in J~m$^{-2}$) for (a) Li-based interfaces and (b) Na-based interfaces. Atomic structures of representative interfaces, namely (c) \ch{Li}(110)$||$\ch{Li2S}(110), (d) \ch{Na}(110)$||$\ch{Na2S}(110) (e) \ch{Li}(100)$||$\ch{Li3P}(001) and (f) \ch{Na}(100)$||$\ch{Na3P}(001). The interface regions are indicated by shaded areas. The non-periodic direction of the interface is indicated by the ``out-of-plane" vectors.
}
\label{fig:interfaces}
\end{figure*}

Figure~\ref{fig:interfaces}a and b show the computed interfacial energetics, $E_f$, $\sigma$, and $W_\mathrm{adhesion}$ (as defined in Eq.~\ref{eq:workadhesion}), for a number of interfaces considered.  An illustration of the interface models for \ch{Li}(110)$||$\ch{Li2S}(110) and \ch{Na}(110)$||$\ch{Na2S}(110) is shown in Figure~\ref{fig:interfaces}c and d, where the interfacial regions are indicated by the shaded areas. Representations of other interfaces are shown in Figure S1-S5 of ESI. In the Li cases considered, we find the most stable interfaces are those formed with \ch{Li3N}, displaying W$_\mathrm{adhesion}$ in the range of 0.8--1.0~J~m$^{-2}$, and $\sigma$ $\sim$0.25~J~m$^{-2}$ (Figure~\ref{fig:interfaces}a). In contrast, the least stable interfaces are \ch{Li}$||$\ch{LiCl}, which exhibit low $W_\mathrm{adhesion}$ and high $\sigma$. In Na-based systems, the most and least stable interfaces are \ch{Na}(110)$||$\ch{Na2O}(110) and \ch{Na}(100)$||$\ch{Na3P}(001), respectively. Note that results of $W_\mathrm{adhesion}$ from Eq.~{\ref{eq:workadhesion2}} (including strain contributions) in Figure S7-S8 appear similar in magnitude (and sign) to those obtained with Eq.~{\ref{eq:workadhesion}} (excluding strain) in Figure~{\ref{fig:interfaces}}. Therefore, we will refer to $W_\mathrm{adhesion}$ of Eq.~{\ref{eq:workadhesion}} and Figure~{\ref{fig:interfaces}} through the remainder of the manuscript.

Previous computational and experimental studies have suggested that Li$||$\ch{Li2S}, Li$||$\ch{Li3P} and Li$||$\ch{LiCl} interfaces are expected to form when argyrodite-\ch{Li6PS5Cl} SE reacts with Li-metal.\cite{wenzelInterfacialReactivityInterphase2018,richardsInterfaceStabilitySolidState2016} A comparison of $W_\mathrm{adhesion}$ (Figure~\ref{fig:interfaces}a) of these interfaces indicates Li$||$\ch{LiCl} $<<$ Li$||$\ch{Li2S} $<$ Li$||$\ch{Li3P}.  Li(100)$||$\ch{Li3P}(001) is expected to dominate the overall interface of Li-metal and \ch{Li6PS5Cl}, if similar quantities of \ch{Li2S} and \ch{Li3P} are produced upon decomposition. In the case of \ch{Li3PS4}, predicted values of  W$_\mathrm{adhesion}$ (Figure~\ref{fig:interfaces}a) suggest the coexistence of both Li$||$\ch{Li2S} and Li$||$\ch{Li3P} interfaces, consistent with prior literature.\cite{richardsInterfaceStabilitySolidState2016,katoXPSSEMAnalysis2018,xiaoUnderstandingInterfaceStability2020} For LiPON, W$_\mathrm{adhesion}$ follows the order \ch{Li}$||$\ch{Li3P}$ << $\ch{Li}$||$\ch{Li2O} $\approx$ \ch{Li}$||$\ch{Li3N}, implying that the Li-metal anode will mostly interface with \ch{Li2O} and \ch{Li3N}, also consistent with  previous investigations.\cite{batesElectricalPropertiesAmorphous1992,schwobelInterfaceReactionsLiPON2015,richardsInterfaceStabilitySolidState2016,chengUnveilingStableNature2020,hoodElucidatingInterfacialStability2021}

In most cases considered, the interfacial region (shaded regions in Figure~\ref{fig:interfaces}c and d) exhibits substantial atomic rearrangements upon  full relaxation, with the exceptions being \ch{Li}(110)$||$\ch{Li3N}(110) (Figure~S3) and  \ch{Li}(110)$||$\ch{LiCl}(100) (Figure~S2). A qualitative analysis of the interface models suggests that there is always a pronounced atomic reconstruction on the metal side of the interface as compared to that of the binary compound for all Li (and Na) interfaces. This is another confirmation that both Li and Na metals are softer than their binary compounds.\cite{dengElasticPropertiesAlkali2016} Li (or Na) atoms originating from the metal side of the interfacial region form stabilising bonds with anion species from the compound side, with bond lengths that are similar to the bulk binary structures (see Table~S5).

In general, interfaces with lattice mismatch smaller than a few percent can be considered as epitaxial, and the re-organization of atoms at the interface remains minimal compared to others with significant lattice mismatch ($\geq 5\%$).  In some cases, we find large lattice mismatches when interfaces are formed from the dominant facets of binary compounds with the (100) or (110) surfaces of the metals (Li or Na). For example, the \ch{Li2S}(111) facet displays a $\sim$14.2\% lattice mismatch with the Li(100) surface (Table~S3), indicating that such an interface may not occur practically. The lattice mismatch between \ch{Li2S}(110) and \ch{Li}(110) facets is lower ($\sim$~5.1\%) and consequently exhibits higher $W_\mathrm{adhesion}$ than \ch{Li2S}(111)$||$\ch{Li}(100). The \ch{Li2S}$||$\ch{Li} interface is likely to exhibit significant structural re-arrangement since \ch{Li2S}(110) facet does not occupy a significant portion of the Wulff volume of \ch{Li2S}, and consequently result in a \ch{Li2S}$||$\ch{Li} interface that is susceptible to delamination in real devices.  

We also find that the surface terminations of binary compounds are crucial to determine the interfacial stability. 
For example, the \ch{Li}(110)$||$\ch{Li2O}(111) interface has a small lattice mismatch of $\sim$1.73\% (Table~S3). However, its fully relaxed geometry exhibits larger lattice distortion of the interfacial region as compared to other Li$||${\ch{Li2O}} based interfaces (see Figure~S4). This interfacial instability comes from the fact that the {\ch{Li2O}}(111) surface is terminated with only Li atoms ---this excess number of Li atoms and lack of anions near the interface region affects the chemical stabilization of the interface due to the lack the bond formation between Li (from the metal side of the interface) and O.  

In Na systems, the Na(110)$||$\ch{Na2S}(110), Na(100)$||$\ch{Na2S}(111), and Na(100)$||$\ch{Na3P}(001) show reconstructions in the interfacial region similar to their Li analogues (Figure~\ref{fig:interfaces}c and Fig. S1c,d). Additionally, we find the computed values of W$_\mathrm{adhesion}$ (and $\sigma$) to be lower (less positive) than their corresponding Li analogues (see Figure~\ref{fig:interfaces}a and b). Despite the low values of W$_\mathrm{adhesion}$ ($<$0.35~J~m$^{-2}$), both Na$||$\ch{Na2S} and Na$||$\ch{Na3P} may still occur at the Na-metal electrode.   The Na$||${\ch{Na2O}} interface has a significantly larger W$_\mathrm{adhesion}$ ($\sim$0.65~J~m$^{-2}$) than Na$||${\ch{Na3P}} ($\sim$0.35~J~m$^{-2}$) and Na$||${\ch{Na2S}} interfaces($\sim$0.30~J~m$^{-2}$).

%%%%%%%%%%%%%%%%%%%%%%%%%%%%%%%%%%%%%%%%%%%%%%%%%%%%%%%%%%%%%%%%%%%%%
\section{Lithium Transport at Heterogeneous Interfaces}
%%%%%%%%%%%%%%%%%%%%%%%%%%%%%%%%%%%%%%%%%%%%%%%%%%%%%%%%%%%%%%%%%%%%%

To quantify ion transport through heterogeneous interfaces, we have used the tracer diffusivity, $D^*$ of Eqs.~\ref{eq:tracer} and \ref{eq:tracer2}. While we quantify only Li-ion transport across heterogeneous interfaces, similar qualitative trends might hold for Na-ion transport as well. 
\begin{eqnarray} 
D^*(T) & = & \lim_{t\rightarrow\infty}{\frac{1}{2dt}}\frac{1}{N}\sum_{i=1}^{N}{\left<{\left|r_i\left(t\right)-r_i\left(0\right)\right|}^2\right>}; \label{eq:tracer}  \\ 
D^*(T) & = & D_0\exp{\left[-\frac{E_a}{k_BT}\right]}. \label{eq:tracer2}
\end{eqnarray}
where $r_i(t)$ is the displacement of the $i^{th}$ Li-ion at time $t$, $N$ is the number of diffusing ions, and $d$ is the dimensionality of the diffusion process. $E_a$ in the Arrhenius Eq.~\ref{eq:tracer2} is the  Li-ion migration energy, $D_0$ is the ionic diffusivity at infinite temperature ($T$), and $k_B$ is the Boltzmann constant. We obtain $D^*$, $D_0$ and $E_a$ from MD simulations based on our trained MTPs,\cite{shapeevMomentTensorPotentials2016} which is machine learned from AIMD simulations of the bulk and interface structures (see  Sec.~\ref{sec:mtpmd}). The largest MD simulations of heterogeneous interfaces investigated in this study contains 8320 atoms and samples the ion dynamics for times  $>$10~ns, which enables an accurate assessment of transport properties. Table S6 summarizes the mean absolute errors from the MTP training and its validation. 

\begin{figure}[!hb]
\includegraphics[width=\columnwidth]{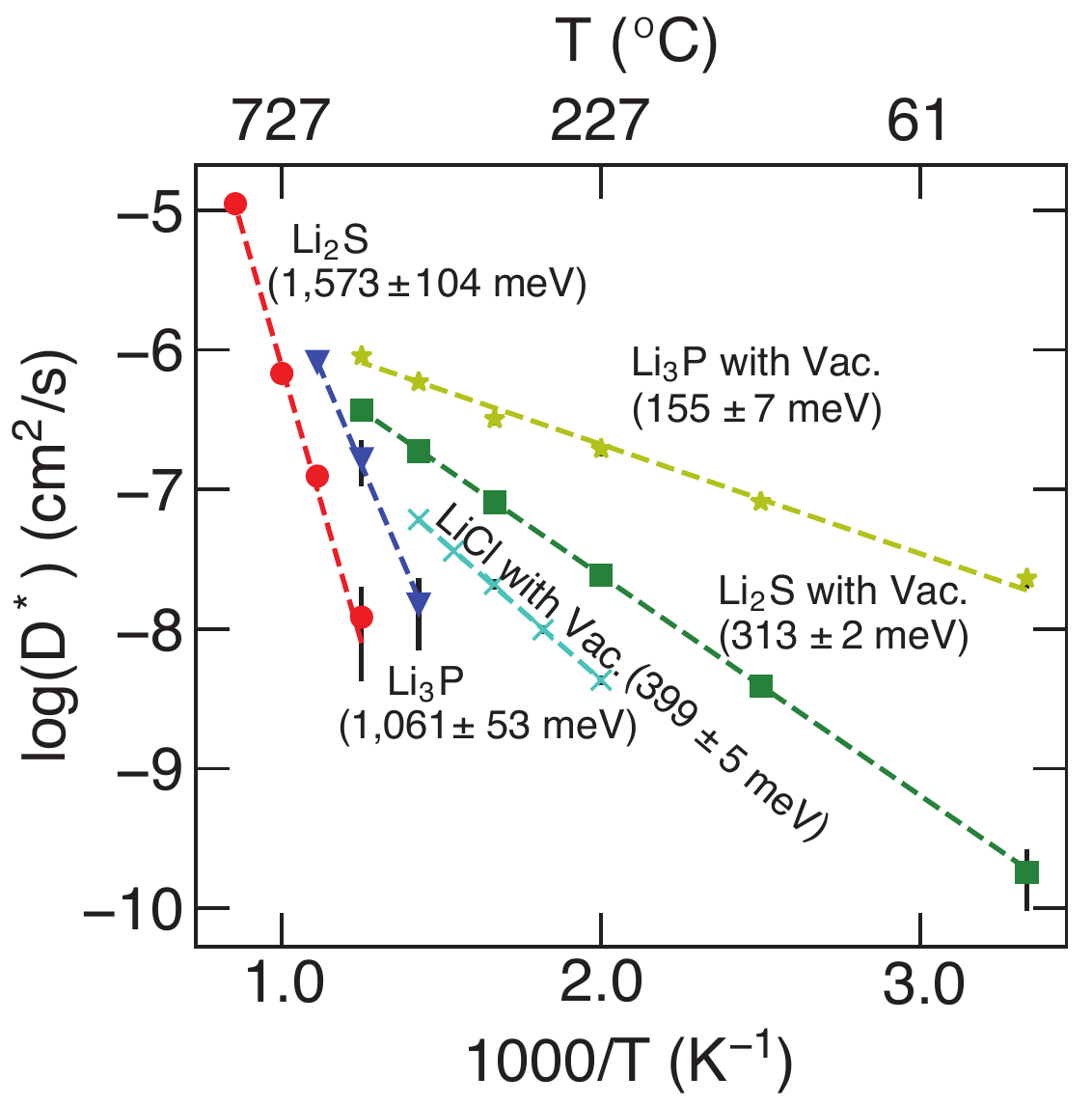}
\caption{
Arrhenius plots of \ch{Li^+} ${D^*}$ (in $\mathrm{cm^{2}~s^{-1}}$) of bulk binary compounds from MTP-MD simulations. The activation energies, calculated from Eq.~\ref{eq:tracer2}, and the related error bars are provided as text annotations. Vac. stands for structures with a Li-vacancy. 
}
\label{fig:singlebinaries}
\end{figure}

The calculated $D^*$ as a function of temperature for bulk binary compounds \ch{Li2S}, \ch{Li3P} and \ch{LiCl}, with and without  \ch{Li^+} vacancies (Vac) are shown in Figure~\ref{fig:singlebinaries}. The assessment of Li-ion transport in the bulk structures of \ch{Li2S}, \ch{Li3P} and \ch{LiCl} is crucial to compare the transport across the heterogeneous interfaces. Notably, our calculated $E_a$ is in reasonable agreement with  experimental results (see Table~S7). For example, the calculated $E_a$ in \ch{LiCl} with Vac, $\sim$399$\pm$5 ~meV), is qualitatively similar to the existing experimental value ($\sim$510~meV\cite{court-castagnetIonicConductivityenhancementLiCl1993}). 
The computed $E_a$ of \ch{Li3P} with Vac ($\sim$155$\pm$7~meV) is in better agreement  with experiment ($\sim$180~meV\cite{nazriPreparationStructureIonic1989}) as compared to the pristine \ch{Li3P} ($\sim$1061$\pm$53~meV). On the other hand, the calculated $E_a$ in pristine-\ch{Li2S} ($\sim$1573$\pm$104~meV) is closer to the experimental value ($\sim$1.5~eV at T~>~800~K\cite{altorferLithiumDiffusionSuperionic1992}) than the calculated $E_a$ in \ch{Li2S} with Vac ($\sim$313$\pm$2~meV). Unsurprisingly, the introduction of Vac lowers the activation energies of both \ch{Li2S} and \ch{Li3P} as shown in Figure~\ref{fig:singlebinaries}. The calculated $E_a$ of  \ch{Li3P} (with Vac) is lower than that of \ch{Li2S} (with Vac), which is in agreement with previous studies showing superior Li-ion conductivity of \ch{Li3P} over \ch{Li2S}.\cite{wangIonicConductionReaction2020}    

\begin{figure*}[ht!]
\includegraphics[width=\textwidth]{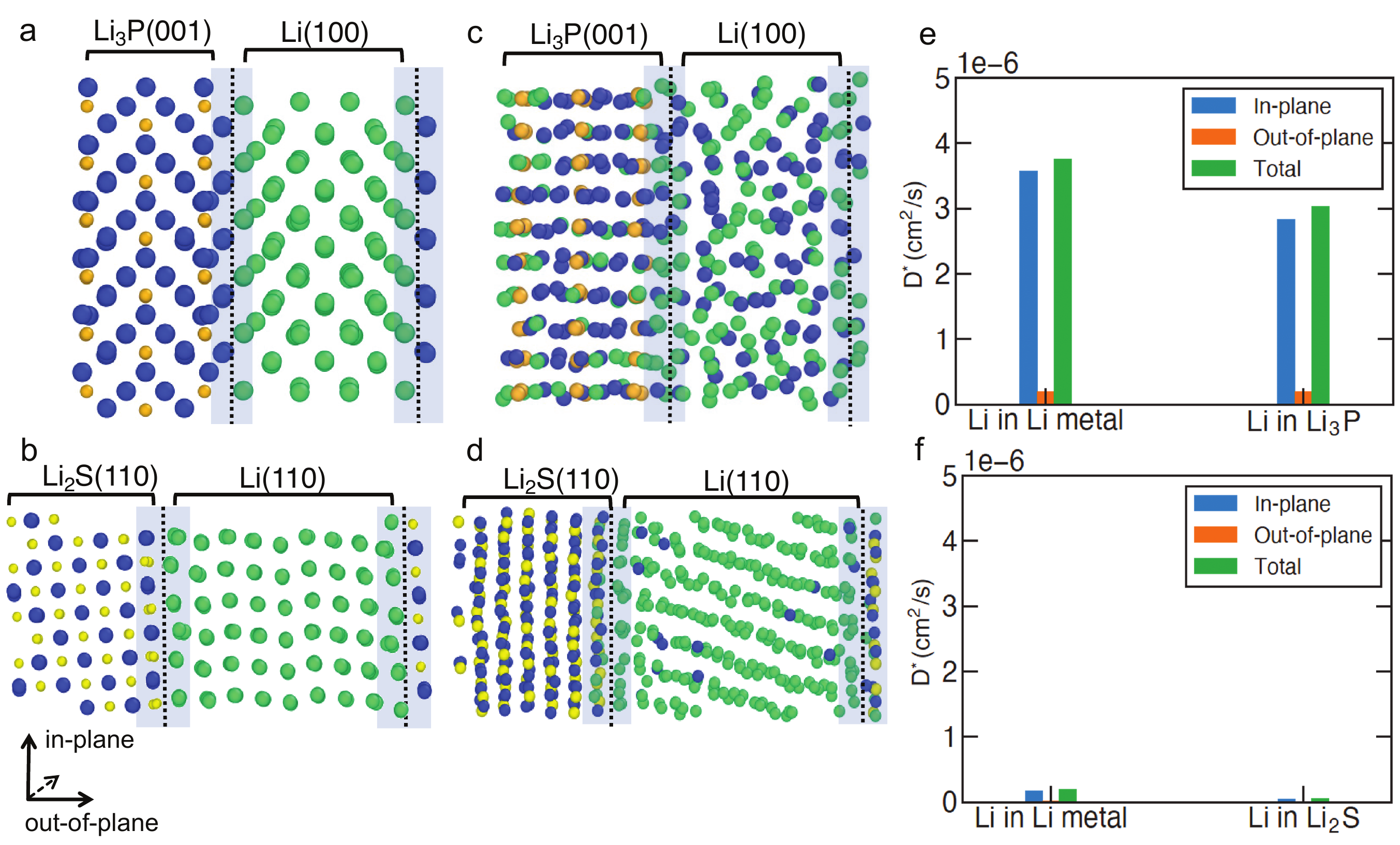}
\caption{Snapshots of (a, c) Li(100)$||$\ch{Li3P}(001) and (b, d) Li(110)$||$\ch{Li2S}(110) interfaces at 0~ns (a and b), and 5~ns (c and d), respectively, at 400 K. The in-plane and out-of-plane components of \ch{Li^+}-$D^*$ in the Li-metal and \ch{Li3P} regions (e) of Li(100)$||$\ch{Li3P}(001) interface and \ch{Li2S} regions (f) of Li(110)\ch{Li2S}(110) interface at 400~K. Dark blue spheres: \ch{Li^+} (binary), green spheres: \ch{Li^+}(metal), orange spheres: P and yellow spheres: S.}
\label{fig:LiLi3PLi2S}
\end{figure*}

To investigate the Li-ion transport across the argyrodite-\ch{Li6PS5Cl}$||$Li-metal interface (i.e., the decomposition products of argyrodite with Li metal), we performed MTP-MD simulations on three interface models, namely, Li(110)$||$\ch{Li2S}(110), Li(100)$||$\ch{Li3P}(001) and Li(110)$||$\ch{LiCl}(100). The choice of these specific interfaces is motivated by their larger $W_\mathrm{adhesion}$ values (Figure~\ref{fig:interfaces}a) compared to other possible configurations using Li-metal and the same binary compound. We randomly introduced a number of \ch{Li^+} vacancies ($\sim$1.1\%) in the interface region to calculate $D^*$, since it is likely that heterogeneous interfaces will comprise highly defective materials, especially due to the \emph{in situ} formed decomposition products. To distinguish \ch{Li^+} belonging either to Li-metal or binary compounds, we have labeled \ch{Li^+} in Li-metal as \ch{Li^+(metal)} (green spheres in Figure~{\ref{fig:LiLi3PLi2S}}, Figure~S9), and \ch{Li^+} in binary compounds as \ch{Li^+(binary)} (dark blue spheres), respectively. Furthermore, the direction of Li-ion transport with respect to the interfacial plane, i.e., in-plane or out-of-plane, helps to qualify the nature of Li transport. Indeed, only Li-ion diffusing out-of-plane will contribute to effective ion-transport across the interface.  The predicted \ch{Li^+}-$D^*$ in both Li-metal and binary compounds are summarized in Table~S8. The mean square displacement (MSD) plots used to derive the \ch{Li^+}-$D^*$ are shown in Figures~S10-S12. 

 %Thus, we trained moment tensor potentials for both, binary compounds \ch{Li2S}, \ch{Li3P} and \ch{LiCl}, and the constructed interfaces Li(110)/\ch{Li2S}(110) (520 atoms), Li(100)/\ch{Li3P}(001) (406 atoms) and Li(110)/LiCl(100) (439 atoms). \juefan{Should we remove this paragraph or move it somewhere else?}

In Figure~\ref{fig:LiLi3PLi2S}a-d and Figure S9a-b, we show the snapshots of different interfaces at 400~K during the MTP-MD simulations. In the following paragraphs, bulk is intended as the portion of the interface model which mimics the bulk structure.
Initially, all interfaces exhibit modest atomic rearrangements near the interface region (violet shaded area). After $\sim$5~ns, significant \ch{Li^+} displacement in both the metal and binary bulk along with \ch{Li^+} exchange (i.e., there is a significant amount of \ch{Li^+}(metal) diffusing into \ch{Li3P} bulk and vice-versa) can be clearly observed in \ch{Li}(100)$||$\ch{Li3P}(001).  This can be understood by  the high values of \ch{Li^+}-$D^*$ (Figure~\ref{fig:LiLi3PLi2S}e) in both the in-plane (within bulk systems, 3.03$\times$10$^{-6}$--3.76$\times$10$^{-6}$ \ch{cm^2}/s)  and out-of-plane (across the bulk systems, 1.93$\times$10$^{-7}$--1.99$\times$10$^{-7}$ \ch{cm^2}/s) directions in the \ch{Li}(100)$||$\ch{Li3P}(001) system. 

In contrast, in \ch{Li}(110)$||$\ch{Li2S}(110) and \ch{Li}(110)$||$\ch{LiCl}(100), we observe limited diffusion events and sparse exchange  of Li-ions during the MTP-MDs, which in turn is quantified by the low in-plane (4.86$\times$10$^{-8}$--1.89$\times$10$^{-7}$ \ch{cm^2}/s) and even lower out-of-plane (6.43$\times$10$^{-9}$--2.21$\times$10$^{-8}$ \ch{cm^2}/s) diffusivities in both systems. We find that for all interfaces, the out-of-plane components of both \ch{Li^+(metal)}  and \ch{Li^+(binary)} are much smaller than their respective in-plane components, which indicate that the \ch{Li^+} diffusion across the interface remains limited.

 %%%%%%%%%%%%%%%%%%%%%%%%%%%%%%%%%%%%%%%%%%%%%%%%%%%%%%%%%%%%%%%%%%%%%
\section{Discussion}
\label{sec:discussion}
%%%%%%%%%%%%%%%%%%%%%%%%%%%%%%%%%%%%%%%%%%%%%%%%%%%%%%%%%%%%%%%%%%%%%

A systematic study of the structures, interfacial energetics, and ionic transport properties of solid/solid interfaces is paramount for the development of solid-state batteries. Here, we have used a combination of accurate DFT calculations to explore the stability of interfaces arising from the decomposition of SEs with highly reducing alkali-metals, i.e., Li and Na. Upon identifying the thermodynamically stable heterogeneous interfaces, we trained MTPs based on accurate AIMD simulations, and in turn used such MTPs to run long duration ($>$10~ns) simulations to elucidate the Li-ion transport properties across specific interfaces.  

Although the morphology of real electrode$||$SE interfaces can be far more complex than the interface models used here, our detailed atomistic models provide insights of the microscopic structure and mechanical stability of buried interfaces between SEs and alkali-metals. Still one major limitation of our analysis is the finite number of interface models considered (20 in this study). Clearly, it is not possible to survey the whole configurational space of interfaces (potentially thousands\cite{haruyamaSpaceChargeLayer2014,butlerDesigningInterfacesEnergy2019,gaoFirstPrinciplesStudyMicroscopic2021}), and alternative strategies should be sought.

This study demonstrates that both surface orientations together with the surface terminations of binary compounds can largely affect the atomistic structures of  interfaces (see Section 3), which in turn determine the interfacial lattice coherence, the thermodynamic stability of interfaces and the mechanical stability of such interfaces in LMBs and solid-state batteries.  Our analysis also suggests that $W_\mathrm{adhesion}$ (of Eq.~\ref{eq:workadhesion}) ---measuring the energy cost to separate two materials of a heterogeneous interface--- is an important descriptor to evaluate the mechanical stability of interfaces.

In particular,  $W_\mathrm{adhesion}$ should be large enough to avoid interface delamination.\cite{wangReviewModelingAnode2018}  Yang et al.\ demonstrated that for common Li-metal$||$SE interfaces, a $W_\mathrm{adhesion}$>~0.7~J~m$^{-2}$ was required to prevent the formation of interfacial voids  with the application of external pressure of 20-30 MPa.\cite{yangInterfacialAtomisticMechanisms2021} Recently, \citeauthor{seymourSuppressingVoidFormation2021}\cite{seymourSuppressingVoidFormation2021} have developed a ''bond breaking'' approach and derived that if $W_\mathrm{adhesion} > 2\gamma$ (where $\gamma$ is the surface energy of Li or Na-metal), the formation of interfacial voids with potential loss of contact  during the Li (or Na) stripping could be avoided. Our data suggest that among Li-based interfaces (Table S4), only the \ch{Li}(100)$||$\ch{Li2O}(110), \ch{Li}(100)$||$\ch{Li3N}(110) and \ch{Li}(110)$||$\ch{Li3N}(110) interfaces satisfy this criterion. For interfaces with Na-metal, only the two Na$||${\ch{Na2O}} interfaces have a $W_\mathrm{adhesion}$ larger than twice of the surface energy of Na(110) (or Na(100)). 

The mechanisms of LiPON passivation of Li-metal has been a matter of debate.\cite{schwobelInterfaceReactionsLiPON2015,hoodElucidatingInterfacialStability2021,chengUnveilingStableNature2020} Recent studies by \citeauthor{hoodElucidatingInterfacialStability2021}\cite{hoodElucidatingInterfacialStability2021} have indicated that \ch{Li3N} and \ch{Li2O} distribute uniformly on the surface of the Li-metal, while \ch{Li3P} was not in direct contact with Li-metal.  In contrast, the study led by the Meng research group  had suggested that only \ch{Li3N}, \ch{Li2O} and \ch{Li3PO4} could be present in the interfacial region formed between Li-metal and LiPON.\cite{chengUnveilingStableNature2020} Our results show that Li$||$\ch{Li2O} and Li$||$\ch{Li3N} interfaces have better interfacial stabilities than Li$||$\ch{Li3P}, which agree well with the experimental scenario that both \ch{Li2O} and \ch{Li3N} can be in direct contact with Li-metal, while \ch{Li3P} can only exist in the sub-interfacial layer.\cite{hoodElucidatingInterfacialStability2021} 

It has been established that argyrodite SEs are prone to decomposition against Li-metal,\cite{wenzelInterfacialReactivityInterphase2018,wangIonicConductionReaction2020} with evidence of formation of \ch{Li2S}, \ch{Li3P} and \ch{LiX} (with \ch{X}=\ch{Cl}, \ch{Br} or \ch{I}) at the potential of Li-metal (i.e., 0 Volts vs.\ \ch{Li/Li^+}). Among the interfaces formed between Li-metal and the decomposition products of argyrodite \ch{Li6PS5Cl} as SE, i.e. Li$||$\ch{Li2S}, Li$||$\ch{Li3P} and Li$||$LiCl, Li$||$\ch{Li3P} has the largest value of $W_\mathrm{adhesion}$ (Figure~\ref{eq:interfaceforamation}), suggesting that \ch{Li3P} is more likely to form a stable interface with Li-metal as compared to the other binary compounds. On one hand, the appreciable electronic conductivity of {\ch{Li3P}} could lead to continuous reactions with Li-metal and growth of the decomposing interphases.\cite{goraiDevilDefectsElectronic2021}  On the other hand, we have not considered the interfacial stability between binary compounds and {\ch{Li6PS5Cl}}. Because  these interfaces may not be mechanically stable, loss of contact between the SE and its decomposition products may also contribute to increased impedance.\cite{wenzelInterfacialReactivityInterphase2018,schwietertClarifyingRelationshipRedox2020}  Indeed, it has been shown that the change in particle size of {\ch{Li2S}} upon lithiation leads to loss of contact of the {\ch{Li6PS5Cl}$||$\ch{Li2S}} interface and increases resistance.\cite{yuUnravellingLiIonTransport2016}

%The majority of existing studies on Li-ion transport are limited to bulk crystalline materials.\cite{wangIonicConductionReaction2020,benitezIonDiffusivitySolid2017,qiBridgingGapSimulated2021} Recently, Wang et al.\cite{wangIonicConductionReaction2020} have used MTP to evaluate the \ch{Li^+} diffusivity of several binary compounds, including  \ch{Li2S}, \ch{Li3P} and \ch{LiCl} that we have also investigated here.  They reported that only \ch{Li3P} among the three exhibits good ionic conductivity.\cite{wangIonicConductionReaction2020} \sai{This paragraph is not needed, no value addition} 

%\st{Here, we have developed the first high-fidelity machine learned potentials (based on moment tensor potential) that enable us to perform large-scale simulations for studying Li-ion transport across multiple Li-metal/SE interfaces. } \sai{Are you trying to say that you don't know whether MTP is transferable or that it is painful to retrain it everytime you have a binary$||$Li interface model? Unclear.} \piero{this is the point. The sentence is perfectly clear...}

The \ch{Li^+} conductivity (or diffusivity) determined in experiments largely depends on the sample quality, its crystallinity and experimental conditions. In particular, the presence of defects, grain boundaries, and lattice disorder all affect the \ch{Li^+} transport significantly.\cite{lacivitaStructuralCompositionalFactors2018,sebtiStackingFaultsAssist2022}  Therefore, here we have restricted our study to the crystalline structures (both decomposing products and interfaces), a situation where the MTP approach has been proven to be adequate to predict ion transport properties.\cite{wangIonicConductionReaction2020, wangLithiumIonConduction2020,qiBridgingGapSimulated2021} However, one major limitation of the current implementation of MTP is its lack of transferability from training within the binary bulk systems to being directly used in heterogeneous interfaces, requiring significant retraining of MTP with new training sets for each distinct interface. Therefore, a complete retraining of the MTP for each interface combination considered in this work is highly resource intensive, which pushes a comprehensive examination of Li (and Na) transport across all interfaces out of the scope of our work. 

Our MTP results suggest that among the three interface models, the Li-metal$||$\ch{Li3P} displays facile Li transport, as shown in Figure~\ref{fig:LiLi3PLi2S}c. However, since only the out-of-plane component of  \ch{Li^+} diffusivity contributes to the active \ch{Li^+} percolation across the interface, these qualitative results show that the interfaces of  Li-metal with \ch{Li3P}, \ch{Li2S} and LiCl are, overall, resistive to Li-ion transport (Figure~\ref{fig:LiLi3PLi2S}, Figure S9) compared to the undecomposed argyrodite SE.\cite{wangIonicConductionReaction2020} 

\begin{figure}[ht!]
\includegraphics[width=\columnwidth]{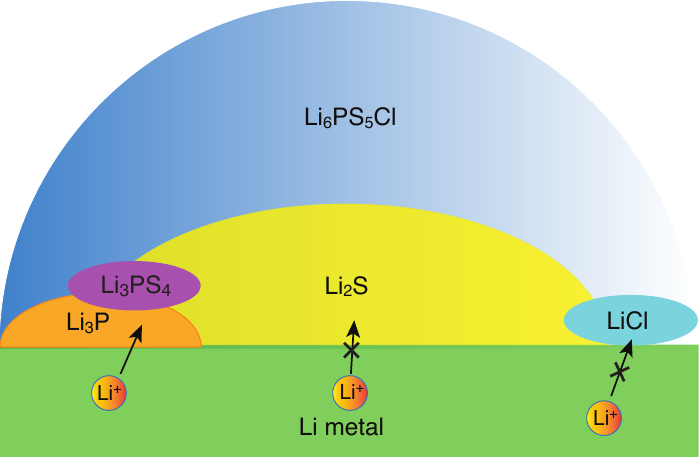}
\caption{
Schematic illustration of a possible structure of the interface between Li-metal and argyrodite-\ch{Li6PS5Cl}, as inferred from the interfacial energetics and \ch{Li}-ion transport simulations. 
}
\label{fig:summary}
\end{figure}

Assume that \ch{Li6PS5Cl} reacts entirely with Li-metal (at 0 Volts vs.\ \ch{Li/Li+}) according to  Eq.~\ref{eq:argyrodite-decomp}: \cite{richardsInterfaceStabilitySolidState2016,wenzelInterfacialReactivityInterphase2018,wangIonicConductionReaction2020}
\begin{equation}
\label{eq:argyrodite-decomp}
    \mathrm{Li_6PS_5Cl} + 8\mathrm{Li}\rightarrow 5\mathrm{Li_2S} + \mathrm{Li_3P} + \mathrm{LiCl}
\end{equation}
where \ch{Li2S} is produced 5$\times$ in excess over the other binaries, in agreement with experimental evidences.\cite{wenzelInterfacialReactivityInterphase2018,schwietertClarifyingRelationshipRedox2020} Moreover, from X-ray photoemission spectroscopy  (XPS) experiments,  \citeauthor{wenzelInterfacialReactivityInterphase2018} \cite{wenzelInterfacialReactivityInterphase2018} and \citeauthor{schwietertClarifyingRelationshipRedox2020}\cite{schwietertClarifyingRelationshipRedox2020} have observed the presence of \ch{Li2S}, \ch{LiCl}, and \ch{Li3P} at the argyrodite$||$Li-metal interface.  On the basis of our interfacial energetics, \ch{Li^+} transport calculations and Eq.~\ref{eq:argyrodite-decomp}, we propose a macroscopic picture of the interface of decomposing argyrodite-\ch{Li6PS5Cl} against Li-metal, as shown in Figure~\ref{fig:summary}.

Our data suggests a lower stability of LiCl$||$Li-metal interface compared to {\ch{Li3P}} and {\ch{Li2S}}, which indicates that LiCl may be in direct contact with Li-metal over a negligible interfacial area. It appears that LiCl may not be directly involved in interfacial Li-transport. At voltages larger than 0.0 Volts vs.\ \ch{Li}/\ch{Li+} other decomposition products have been reported and observed, with the most prominent being \ch{Li3PS4},\cite{richardsInterfaceStabilitySolidState2016,zhuFirstPrinciplesStudy2016,schwietertClarifyingRelationshipRedox2020} which may form in the sub-interfacial layers of the SE. Furthermore, our MTP-MD demonstrated that Li percolation in the Li$||$\ch{Li3P} interface is facile compared to the other interfaces as signified by the black arrows in Figure~\ref{fig:summary}.

%%%%%%%%%%%%%%%%%%%%%%%%%%%%%%%%%%%%%%%%%%%%%%%%%%%%%%%%%%%%%%%%%%%%%
\section{Conclusion}
%%%%%%%%%%%%%%%%%%%%%%%%%%%%%%%%%%%%%%%%%%%%%%%%%%%%%%%%%%%%%%%%%%%%%
Chalcogen-containing SEs show among the highest room temperature ionic conductivities  ($\sim$10$^{-2}$~S~cm$^{-1}$), but their practical applications in LMBs are limited by the decomposing interfaces when in contact with Li metal. Similar constraints bottleneck the implementation of SEs in NIBs as well. Therefore, it is vital to understand the interfacial properties of these decomposing interfaces, either experimentally or theoretically. In this work, we have systematically evaluated the thermodynamic stability (of Li- and Na-systems) and Li-ion transport properties of multiple decomposing interfaces, by employing first-principles calculations and large-scale MD simulations based on MLPs.  Our results reveal that the interfacial stability of decomposition products with alkali-metals is largely affected by the surface properties of the decomposition products. In general, we have observed that the interfaces formed between alkali-metal with argyrodite-\ch{Li6PS5Cl} are resistive, to Li-ion transport. Finally, our high-fidelity MLPs, trained explicitly for interfaces, shed light on the complicated interfacial transport properties, which will aid in the study and optimization of SEs in the future.

%%%%%%%%%%%%%%%%%%%%%%%%%%%%%%%%%%%%%%%%%%%%%%%%%%%%%%%%%%%%%%%%%%%%
\section{Methods}
\label{sec:methods}
%%%%%%%%%%%%%%%%%%%%%%%%%%%%%%%%%%%%%%%%%%%%%%%%%%%%%%%%%%%%%%%%%%%%%
\subsection{First-principles Calculations}
DFT was used to approximate the energy contributions introduced in Sec.~\ref{sec:quantities}. The wavefunctions were described using  plane-waves for the valence electrons together with projected augmented wave potentials for the core electrons as implemented in the Vienna Ab-initio Simulation Package (VASP).\cite{blochlProjectorAugmentedwaveMethod1994,kresseUltrasoftPseudopotentialsProjector1999,kresseEfficientIterativeSchemes1996} The exchange-correlation contributions were treated within the generalized gradient approximation (GGA) as parameterized by Perdew, Burke, and Ernzerhof (PBE).\cite{perdewGeneralizedGradientApproximation1996} The valence electron configurations for each element were as follows: \ch{Li:\, s^1p^0}, \ch{N:\, s^2p^3}, \ch{O:\, s^2p^4}, \ch{Na:\, s^1p^0}, \ch{P:\, s^2p^3}, \ch{S:\, s^2p^4} and \ch{Cl:\, s^2p^5}.  The parameters we used for geometry optimization, surface energy and interfacial energetics calculations of the binary compounds and the constructed interfaces follow the MITRelaxSet, as  in pymatgen.\cite{ongPythonMaterialsGenomics2013}  We used a plane wave energy cutoff of 520~eV and a $k$-point mesh generated using a $k$-point density of 25 \AA$^{-1}$. The total energy of each structure was converged to 10$^{-5}$ eV/cell, and the geometry optimizations were stopped when the change in total energy was smaller than 10$^{-4}$~eV between two subsequent ionic steps.

AIMD were performed with VASP to generate the initial training sets for the MTP-MD (see Sec.~\ref{sec:mtpmd}). A plane-wave energy cutoff of 400 eV and a $\Gamma$-only $k$-mesh were used. The canonical ensemble (NVT) was achieved using Nos\'{e}-Hoover thermostat and  a time step of 2 fs.\cite{noseUnifiedFormulationConstant1984,hooverCanonicalDynamicsEquilibrium1985} Since previous studies have reported\cite{novoselovMomentTensorPotentials2019,wangLithiumIonConduction2020} that the training set for MTP-MD should cover the whole configurational space and contain sufficient data so as to rarely invoke DFT calculations, we performed AIMD calculations at 1000~K for 14-20~ps (preceded by a temperature ramping of 2~ps), which resulted in training sets containing 7000-10000 configurations. The supercell sizes used for binary compounds pristine structures were 4$\times$4$\times$4 for \ch{Li} metal (128 atoms), 2$\times$2$\times$2 for \ch{Li2S} (96 atoms), 3$\times$3$\times$3 for \ch{Li3P} (216 atoms) and 3$\times$3$\times$3 for \ch{LiCl} (216 atoms). We also studied the vacancy-mediated diffusion by creating Li$^+$ vacancies inside the Li metal and binary compounds. 

Li$^+$ vacancies were introduced by removing Li atoms and compensating with a uniform (jellium) charge background. Also, we created specific supercells that enabled a Li$^+$ vacancy concentration of $\sim$0.8\% for all compounds, which can arise at a synthesis temperature of 1200~K with a defect formation energy of 0.5~eV. Specifically, we used supercells of 4$\times$4$\times$4 with one \ch{Li^+} vacancy for \ch{Li} metal (127 atoms), 2$\times$2$\times$4 with one \ch{Li^+} vacancy for \ch{Li2S} (191 atoms), 3$\times$3$\times$3 with one \ch{Li^+} vacancy for \ch{Li3P} (215 atoms) and 3$\times$3$\times$3 with one \ch{Li^+} vacancy for \ch{LiCl} (215 atoms). To study the \ch{Li^+} transport across \ch{Li}-metal$||$decomposition product interfaces, we have created \ch{Li^+} vacancies randomly in the interface region (shaded regions in Figures~{\ref{fig:interfaces}} and {\ref{fig:LiLi3PLi2S}}, with a vacancy concentration of $\sim$1.1\%. The interfaces that we chose were \ch{Li}(110)$||$\ch{Li2S}(110) (520 atoms), \ch{Li}(100)$||$\ch{Li3P}(001) (406 atoms) and \ch{Li}(110)$||$\ch{LiCl}(100) (439 atoms). 

%%%%%%%%%%%%%%%%%%%%%%%%%%%%%%%%%%%%%%%%%%%%%%%%%%%%%%%%%%%%%%%%%%%%%
\subsection{Moment-Tensor Potential Molecular Dynamics}
\label{sec:mtpmd}
%%%%%%%%%%%%%%%%%%%%%%%%%%%%%%%%%%%%%%%%%%%%%%%%%%%%%%%%%%%%%%%%%%%%%
MTP for bulk and interfaces investigated in this study were trained using the machine learning of interatomic potentials (MLIP) package.\cite{podryabinkinActiveLearningLinearly2017}  In the training of the MTP potentials, several parameters need to be carefully selected to  balance  computational cost vs.~accuracy of the trained potentials. During training, we have extensively tested the effects of weights on reproducing the \emph{ab initio} total energies, forces and stresses, as well as cutoff radius (R$\mathrm{cut}$) and the maximum level of basis functions (lev$_\mathrm{max}$) on the accuracy of energy and forces of trained MTP potentials. We concluded that a ratio of weights of 100:10:1 for energies, forces, and stresses, respectively, was appropriate to achieve good accuracy. Also, we found that a lev$_\mathrm{max}$ of 10 and a R$\mathrm{cut}$ of 5~\AA, provided a tolerable level of fitting and validation errors in energies(< 10 meV) and forces (< 30 meV/\AA), as documented in Table~S6.

Since our MTPs were trained at high temperatures ($\sim$1000~K), we further validated the transferability of the potentials to lower temperatures (i.e., 300-500~K). Specifically, we constructed validation sets by performing AIMD at 300~K/500~K for 4~ps ($\sim$2000 snapshots for each temperature). The fitting and validation errors on the total energies in both binary compounds and interface models were always $<$~10~meV, while the errors on forces were within $\sim$30~meV~\AA$^{-1}$. 

Upon training, MTP-MD were performed using LAMMPS,\cite{plimptonFastParallelAlgorithms1995} where the MD simulations were performed in the temperature range of 300-1000~K at intervals of 100~K. A Nos\'{e}-Hoover thermostat was used to simulate the canonical ensemble (NVT).\cite{noseUnifiedFormulationConstant1984,hooverCanonicalDynamicsEquilibrium1985} Long MD simulations were carried out for at least 10~ns with a short timestep of 1~fs, preceded by a temperature ramping for 100~ps and an equilibration period of 1~ns to reach each target temperature. We also benchmarked our MTP $D^*$ data with AIMD results (see Table S9). Specifically, we find that our MTP-MD calculated $D^*$ at 900~K and 800~K are in reasonable agreement with AIMD calculations at the same temperatures, signifying the high fidelity of our MTP-MD simulations.

\subsection{Validation of Interfacial Models}
To verify the accuracy of our methodology in predicting interfacial properties, we have calculated interfacial energetics using two additional ``constrained'' optimization methods, namely, \emph{i}) ``Fix-binary": the middle layers of the decomposition product was fixed to mimic the bulk in-plane lattice constants of the binary compound, and, \emph{ii}) ``Fix-metal": middle layers of Li(Na) metal are fixed. The default method used throughout the work is when we do not constrain the middle layers of either binary compounds or the metal, referred to as ``Fully-relaxed". To test these scenarios, we chose \ch{Na}(110)$||$\ch{Na2O}(110) for Na-based and \ch{Li}(100)$||$\ch{Li3P}(001) for Li-based interfaces, respectively. The calculated $E_f$, with and without constrained optimization, are shown in Figure~S6. Notably, $E_f$ calculated using constrained optimization is $\sim$0.02~J~m$^{-2}$ and $\sim$0.1~J~m$^{-2}$ higher than Fully-relaxed for \ch{Li}(100)$||$\ch{Li3P}(001), and \ch{Na}(110)$||$\ch{Na2O}(110), respectively.

%%%%%%%%%%%%%%%%%%%%%%%%%%%%%%%%%%%%%%%%%%%%%%%%%%%%%%%%%%%%%%%%%%%%%%
%The "Acknowledgement" section can be given in all manuscript % classes.
%This should be given within the "acknowledgement" % environment, which
%will make the correct section or running title.
%%%%%%%%%%%%%%%%%%%%%%%%%%%%%%%%%%%%%%%%%%%%%%%%%%%%%%%%%%%%%%%%%%%%%

\begin{acknowledgments}
 J.W.\, A.A.P. and P.C.\ acknowledges funding
from the National Research Foundation under his NRF Fellowship
NRFF12-2020-0012. The
computational work was performed on resources of the National
Supercomputing Centre, Singapore (\url{https://www.nscc.sg}). 
\end{acknowledgments}

%%%%%%%%%%%%%%%%%%%%%%%%%%%%%%%%%%%%%%%%%%%%%%%%%%%%%%%%%%%%%%%%%%%%%%
%The appropriate \bibliography command should be placed here.  % Notice
%that the class file automatically sets \bibliographystyle % and also
%names the section correctly.
%%%%%%%%%%%%%%%%%%%%%%%%%%%%%%%%%%%%%%%%%%%%%%%%%%%%%%%%%%%%%%%%%%%%%
\bibliography{biblio}

\end{document}